\RequirePackage{fix-cm}
\documentclass[smallextended]{svjour3}       
\smartqed  
\usepackage{graphicx}
\begin{document}

\title{Bending back stress chains and unique behaviour of granular matter in cylindrical geometries
}


\author{Raphael Blumenfeld   \and   Julian Ma 
}


\institute{R. Blumenfeld \at
1. Imperial College, London, UK\\
2. College of Science, NUDT, Changsha, Hunan, China\\
3. Cavendish Laboratory, Cambridge, UK\\
\email{rbb11@cam.ac.uk}           
           \and
J. Ma \at
1. College of Science, NUDT, Changsha, Hunan, China\\
2. Cavendish Laboratory, Cambridge, UK\\
}

\date{Received: date / Accepted: date}

\maketitle

\begin{abstract}
We analyse the general solutions for the stress field in planar annuli of isostatic media, a model often used for marginally rigid granular materials in Couette cells.
We demonstrate that these solutions are much richer than in rectangular symmetries. 
Even for uniform media, stress chains are found to curve, broaden away from the stress source, attenuate and leak stress into a cone of influence. Most spectacularly, stress chains may bend back and transmit forces oppositely to the original direction. None of these phenomena arises in solutions for uniform media in Cartesian coordinates. 
We further analyse non-uniform media, which exhibit chain branching and stress leakage from the chains.
These results are directly relevant to the many experiments on granular materials, carried out in Couette cells. They also shed light on, and are supported by, hitherto unexplained experimental observations of curved and back-bending chains, which we point out. 
In particular, we use our results to provide a new interpretation for the pattern of slip lines observed experimentally.
\end{abstract}

\section{Introduction}
\label{intro}
The ubiquity and significance of granular matter (GM) have focused scientific and technological attention for millenia \cite{Bletal15}, but the theoretical understanding of this form of matter is far from complete. One of the most important theoretical challenges is the development of a theory of stress transmission in dry such media. Predicting stress is essential to a wide range of technological and geo-mechanical applications, as well as being a springboard for modelling the dynamic behaviour of GM. 
Experimental observations of nonuniform stress transmission in GM, e.g. via 'force chains', date back to the 1940s \cite{Seelig46,EarlyChains}, with modern experiments revealing detailed features of this phenomenon \cite{LateChains}. 
The limitations of conventional theories, such as elasticity, to account for force chains led to investigations of ideal systems: isostatic media. These are statically determinate and marginally rigid aggregates, characterised by a low mean contact (or coordination) number, whose intergranular forces can be determined, in principle, from balance conditions alone. The stress equations of isostaticity theory (IT) are hyperbolic, differing markedly from the conventional elliptic equation of elasticity theory \cite{EdOa89,EdGr99,Edinburgh,BaBl02,Bl04}.  
Understanding these ideal systems is an essential step to a theory of real GM, which comprise both isostatic regions and denser regions, where conventional theories are valid \cite{Bl04}. 

Many physical and numerical experiments are carried out in cylindrical setups \cite{CouetteBasedWork}, yet most theoretical analyses of stresses in isostatic materials are based on rectangular coordinate systems. This practice is based on an implicit, normally little discussed, assumption that the stress chains phenomenon is independent of the system symmetry. The purpose of this paper is to show that, at least in two dimensions (2D), this is not the case, namely, that isostatic stress solutions in cylindrical symmetry have features that do not arise in rectangular coordinates. This we do by solving explicitly for the stress field in annuli and then demonstrating these solutions by examples.
Specifically, we show that uniform isostatic media exhibit stress chains that may curve, broaden, dissipate and even bend backwards. The characteristics of these solutions are analysed and their implications discussed. Experimental support in the literature for some of these phenomena are pointed out and interpreted in view of these solutions.

\section{The isostaticity stress equations}
\label{sec:2}
In 2D Cartesian coordinates, the isostaticity stress field equations are \cite{Edinburgh,BaBl02}
\begin{eqnarray}
\vec{\nabla}\sigma & = & \vec{g} \label{eq:Balance1} \\
\sigma & = & \sigma^T \label{eq:Balance2} \\
p_{xx}\sigma_{yy} & - & 2p_{xy}\sigma_{xy} + p_{yy}\sigma_{xx} = 0 \label{eq:ConstEq} \ ,
\end{eqnarray}
where $\sigma$ is the stress tensor, $\vec{g}$ includes all external and body forces and $p_{ij}$ are the components of a fabric tensor, $\mathcal{P}(x,y)$, which can be determined directly from the grain structure \cite{BaBl02} and upscaled to the continuum \cite{Bl04a}. 
It has been shown that the determinant of $\mathcal{P}$ is generically negative \cite{Bl04} and therefore that these equations are hyperbolic and yield stress chain solutions, \cite{Edinburgh,BaBl02}. Equations (\ref{eq:Balance1})-(\ref{eq:ConstEq}) have been analysed and solved for uniform fabric tensors \cite{Edinburgh,EdGr99,Bl04}, as well as for position dependent ones \cite{Geetal08}. Yet, in spite of numerous experiments in annuli, e.g. Couette cells, exact solutions in the literature are often derived in rectangular coordinates, presumably under the assumption that the behaviour would not differ significantly in cylindrical geometries. This is not the case. 
Moreover, since force chains often originate in very localised regions, down to single grains, it is more appropriate to describe the field around such an origin in polar coordinates.

In polar coordinates, the balance equations are
\begin{eqnarray}
\partial_r\left(r\sigma_{rr}\right) & + & \partial_\theta\sigma_{r\theta} - \sigma_{\theta\theta} = g_r \label{eq:Balance3} \\
\partial_r\left(r\sigma_{r\theta}\right) & + & \partial_\theta\sigma_{\theta\theta} + \sigma_{r\theta} = g_\theta \label{eq:Balance4} \ .
\end{eqnarray}
The equations are closed by a stress-structure relation that, for consistency, has the same form as (\ref{eq:ConstEq}): 
\begin{equation}
\pi_{rr}\sigma_{\theta\theta} - 2\pi_{r\theta}\sigma_{r\theta} + \pi_{\theta\theta}\sigma_{rr} = 0 \ .
\label{eq:ConstEq2}
\end{equation}

\section{Analysis}
\label{sec:3}
Rewriting the stress components of equation (\ref{eq:ConstEq2}) in Cartesian coordinates and comparing to (\ref{eq:ConstEq}), a relation can be derived between the Cartesian, $\mathcal{P}$, and polar, $\Pi$, fabric tensors:
\begin{equation}
\left( \begin{array}{c} p_{xx} \cr p_{yy} \cr p_{xy} \end{array} \right) =
\left( \begin{array}{ccc} 1 & -C & -S \cr 1 & C & S \cr 0 & -\frac{S}{2} & C \end{array} \right)
\left( \begin{array}{c} \pi_{rr} + \pi_{\theta\theta}\cr  \pi_{rr} - \pi_{\theta\theta} \cr \pi_{r\theta} \end{array} \right) \ ,
\label{eq:PPTransform}
\end{equation}
where, for brevity, $(S,C)\equiv(\sin{2\theta},\cos{2\theta})$. 
The condition that $det\{\mathcal{P}\}<0$ translates to:
\begin{eqnarray}
\frac{3S^2}{4}(\pi^2_{rr} + \pi^2_{\theta\theta}) & + & (2-\frac{3S^2}{4})\pi_{rr}\pi_{\theta\theta} - \nonumber \\
- CS(\pi_{rr} - \pi_{\theta\theta})\pi_{r\theta} & - & \pi^2_{r\theta} < 0  \ .
\label{eq:PiNegDet}
\end{eqnarray}
From this condition we can determine a region in the constitutive parameter space $\pi_{rr}$-$\pi_{\theta\theta}$-$\pi_{r\theta}$, in which the determinant of $\Pi$ is negative {\it for all values of} $\theta$.  Note that this eliminates a wide range of fabric tensors, some of which were studied in the literature, e.g. in \cite{Edinburgh,Geetal08}.

To obtain the general solution for the stress, we follow a similar procedure to the one initiated in \cite{Geetal08}. 
Assuming first that  $\pi_{\theta\theta}\neq 0$, we define $q_{ij}\equiv\pi_{ij}/\pi_{\theta\theta}$ and 
substitute $\sigma_{rr}$ from the stress-structure condition (\ref{eq:ConstEq2}) into equation (\ref{eq:Balance3}). The resulting balance equations can be written as

\begin{equation}
A\partial_\rho \vec{u} + \partial_\theta\left(\vec{u}\right) - B\vec{u} = \vec{g}\ ,
\label{eq:Balance5}
\end{equation}
with $\rho\equiv\ln{r/r_0}$, $\vec{u} \equiv \left(\sigma_{r\theta},\sigma_{\theta\theta}\right)$, $B\equiv\left( \begin{array}{cc} -2\left(1+\partial_\rho\right)q_{r\theta} & \left[1+ \left(1+\partial_\rho\right)q_{rr}\right] \cr -2 & 0 \end{array} \right)$ and $A\equiv \left( \begin{array}{cc} 2q_{r\theta} & -q_{rr} \cr 1 & 0 \end{array} \right)$.
The characteristic variables, $w_{1,2}$, can be expressed in terms of the stress components as
\begin{equation}
\vec{u} = Y\vec{w} = \left( \begin{array}{cc} 1 & 1 \cr \lambda_1 & \lambda_2 \end{array} \right) \left( \begin{array}{c} w_1 \cr w_2 \end{array} \right)\ ,
\label{eq:Chars}
\end{equation}
where $\lambda_{1,2} = q_{r\theta} \pm \sqrt{q^2_{r\theta}-q_{rr}}$ are the eigenvalues of $A$ and $Y^{-1}AY=\Lambda$ is diagonal. 
Since $q^2_{r\theta}-q_{rr}=-det\left\{\Pi\right\}/\pi^2_{\theta\theta}>0$, $\lambda_{1,2}$ are real and distinct.
In terms of $\vec{w}$, (\ref{eq:Balance5}) becomes
\begin{equation}
\Lambda\partial_\rho \vec{w} + \partial_\theta \vec{w} = Y^{-1}\left(BY-A\partial_\rho Y-\partial_\theta Y\right)\vec{w} + \vec{h} \ ,
\label{eq:Eigens}
\end{equation}
where $\vec{h}=Y^{-1}\vec{g}$.

It is now convenient to parameterise the characteristic paths by length variables, $s_i$: $\partial_\rho s_i = 1/\lambda_i$ and $\partial_\theta s_i = 1$ ($i = 1,2$). For spatially uniform $\Pi$, $\nabla \pi_{ij}=0$, this reduces (\ref{eq:Eigens}) to the linear form
\begin{equation}
\frac{d\vec{w}}{ds} = Y^{-1}BY\vec{w} + \vec{h} \ .
\label{eq:Eigens1}
\end{equation}
This equation shows that, even for spatially uniform fabric tensors, $w_1$ and $w_2$ are coupled by the off-diagonal terms of $Y^{-1}BY$. This is in contrast to the situation in rectangular coordinates, where uniform fabric tensors give rise to {\it decoupled} characteristics and to straight path solutions, on which the stress is constant whilst vanishing elsewhere. The inherent coupling in polar coordinates makes stress chains in annuli a much richer phenomenon, which we proceed to explore.

\begin{figure}[h]
\begin{center}
\includegraphics[width=5cm]{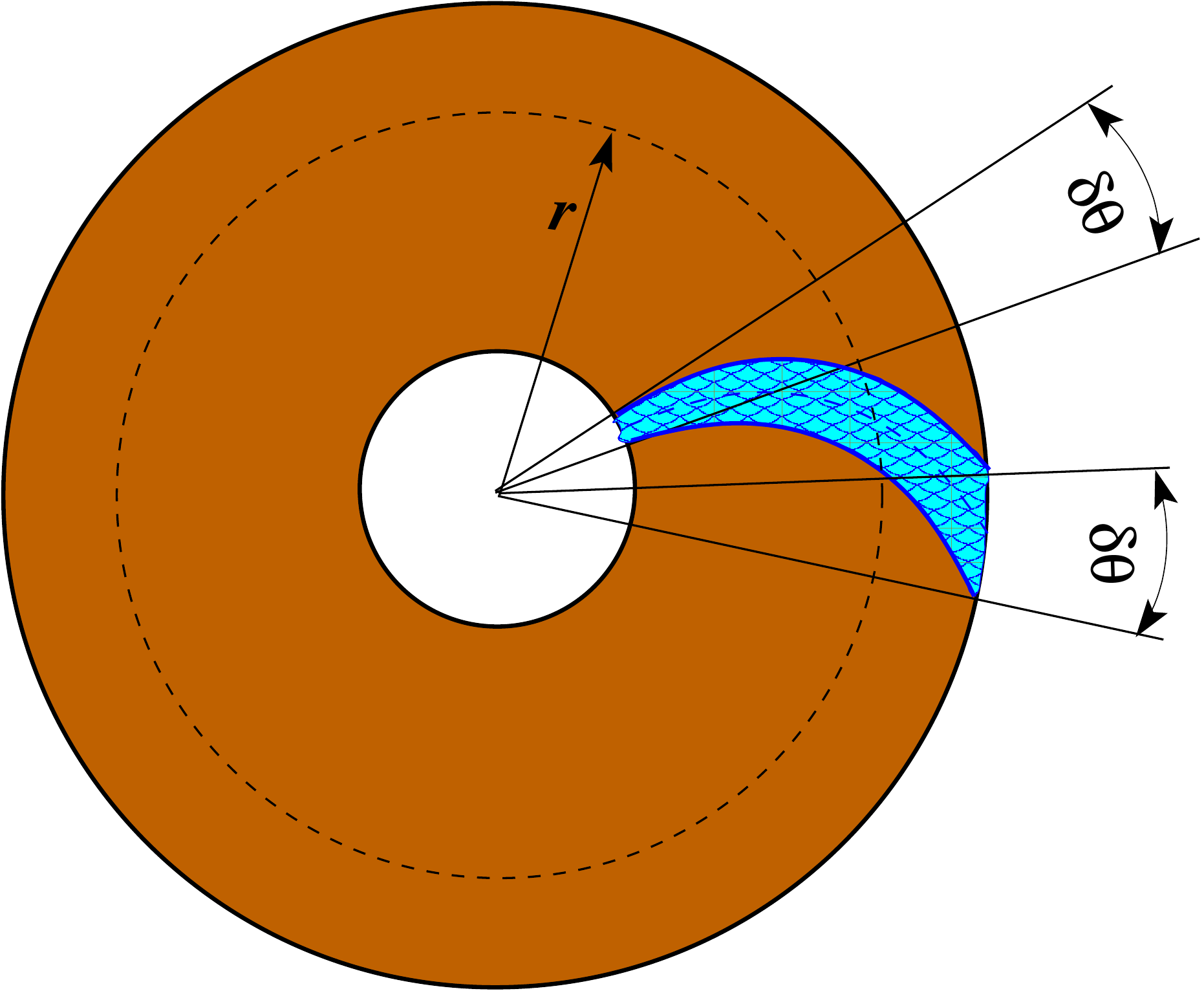}%
\caption{\label{fig:ChainFlare} 
The right hand side characteristic path (in blue), emanating from a source of width $r_0\delta\theta$ at the internal boundary, flares out and broadens to $r\delta\theta$.}
\end{center}
\end{figure}

Consider the response to a localised stress source at the inner boundary, $r_0$. Clearly, the linearity of the equations guarantees that the response to any stress source distribution can be found by superposition.
Along the characteristic paths, we have $\partial_\rho \theta = 1/\lambda_i$. Therefore, a point source at $(r,\theta)=(r_0,\theta_0)$ gives rise to a characteristic path, whose trajectory satisfies
\begin{equation}
\theta_i-\theta_0=\frac{1}{\lambda_i} \ln \left(r/r_{0}\right) \ .
\label{eq:ChainFlare}
\end{equation}
It follows that a stress source of width $r_0\delta\theta$ at the internal boundary generates two stress chains whose trajectories `flare out' and broaden.This is illustrated in figure \ref{fig:ChainFlare} for the right hand side characteristic path.

For completeness, we present an alternative analysis that allows $\pi_{\theta\theta}=0$. In this case, $\pi_{rr}\neq 0$ ($\pi_{\theta\theta}$ and $\pi_{rr}$ cannot both vanish lest the equations are no longer hyperbolic). Following the same procedure as above, we now scale the fabric tensor, $\tilde{q}_{ij}\equiv\pi_{ij}/\pi_{rr}$, substitute from relation (\ref{eq:ConstEq2}) into (\ref{eq:Balance3})--(\ref{eq:Balance4}) and rewrite these as
\begin{equation}
\partial_\rho \vec{v} + \partial_\theta\left( \tilde{A}\vec{v}\right) - \tilde{B}\vec{v} = \vec{g}\ ,
\label{eq:Balance5a}
\end{equation}
where 
$\tilde{A}\equiv\left( \begin{array}{cc} 0 & 1 \cr -\tilde{q}_{\theta\theta} & 2\tilde{q}_{r\theta} \end{array} \right)$, 
$\tilde{B}\equiv\left( \begin{array}{cc} -1-\tilde{q}_{\theta\theta} & 2\tilde{q}_{r\theta} \cr 0 & -2 \end{array} \right)$ and 
$\vec{v} \equiv \left(\sigma_{rr}, \sigma_{r\theta}\right)$.
In terms of $\vec{v}$, the characteristic variables, $\tilde{w}_{1,2}$, are now
\begin{equation}
\vec{v} = \tilde{Y}\vec{\tilde{w}} = \left( \begin{array}{cc} 1 & 1 \cr \tilde\lambda_1 & \tilde\lambda_2 \end{array} \right) \left( \begin{array}{c} \tilde{w}_1 \cr \tilde{w}_2 \end{array} \right)\ ,
\label{eq:Charsa}
\end{equation}
where $\tilde\lambda_{1,2} = \tilde{q}_{r\theta} \pm \sqrt{\tilde{q}^2_{r\theta}-\tilde{q}_{\theta\theta}}$ are the eigenvalues of $\tilde{A}$ and $\tilde{Y}^{-1}\tilde{A}\tilde{Y}=\tilde\Lambda$ is diagonal. 
Since $\tilde{q}^2_{r\theta}-\tilde{q}_{\theta\theta}=-det\left\{\Pi\right\}/\pi^2_{rr}>0$, these eigenvalues are also real and distinct.
In terms of $\vec{\tilde{w}}$, (\ref{eq:Balance5a}) becomes
\begin{equation}
\partial_\rho \vec{\tilde{w}} + \tilde\Lambda\partial_\theta \vec{\tilde{w}} = \tilde{Y}^{-1}\left( \tilde{B}\tilde{Y} - \partial_\rho\tilde{Y} - \tilde{A}\partial_\theta \tilde{Y} \right)\vec{\tilde{w}} + \vec{\tilde{h}} \ ,
\label{eq:Eigensa}
\end{equation}
where $\vec{\tilde{h}}\equiv\tilde{Y}^{-1}\vec{g}$. Unsurprisingly, the forms of equations (\ref{eq:Eigensa}) and (\ref{eq:Eigens}) are very similar. 
Again, we see that the characteristics are coupled by $\tilde{Y}^{-1}\tilde{B}\tilde{Y}$ even for spatially uniform fabric tensors. The length parameters along the characteristic paths, $s_i$, are now: $\partial_\rho s_i = 1$ and $\partial_\theta s_i = 1/\tilde\lambda_i$ ($i = 1,2$). Along the paths $\partial_\rho \theta_i = \tilde\lambda_i$ and the two paths flare out following the relation 

\begin{equation}
\theta_i=\tilde\lambda_i \ln \left(r/r_{0}\right) \ .
\label{eq:ChainFlarea}
\end{equation}
For spatially uniform fabric tensors, equation (\ref{eq:Eigensa}) reduces to 
\begin{equation}
\frac{d\vec{\tilde{w}}}{ds} = \tilde{Y}^{-1}\tilde{B}\tilde{Y}\vec{\tilde{w}} + \vec{\tilde{h}} \ .
\label{eq:Eigens1a}
\end{equation}

\section{Example solutions}
\label{sec:4}
\begin{figure}[h]
\begin{center}
\includegraphics[width=5cm]{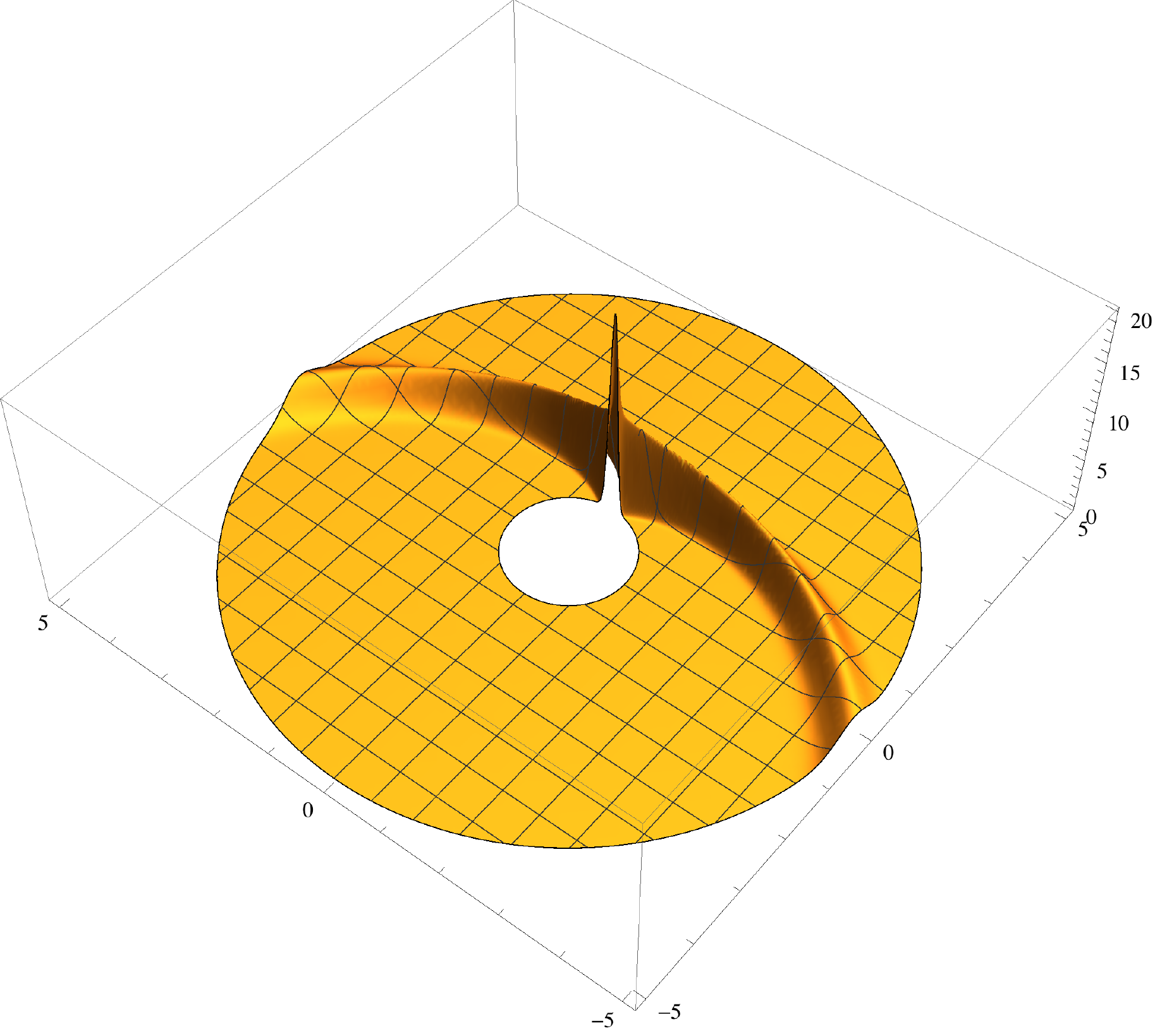}%
\includegraphics[width=5cm]{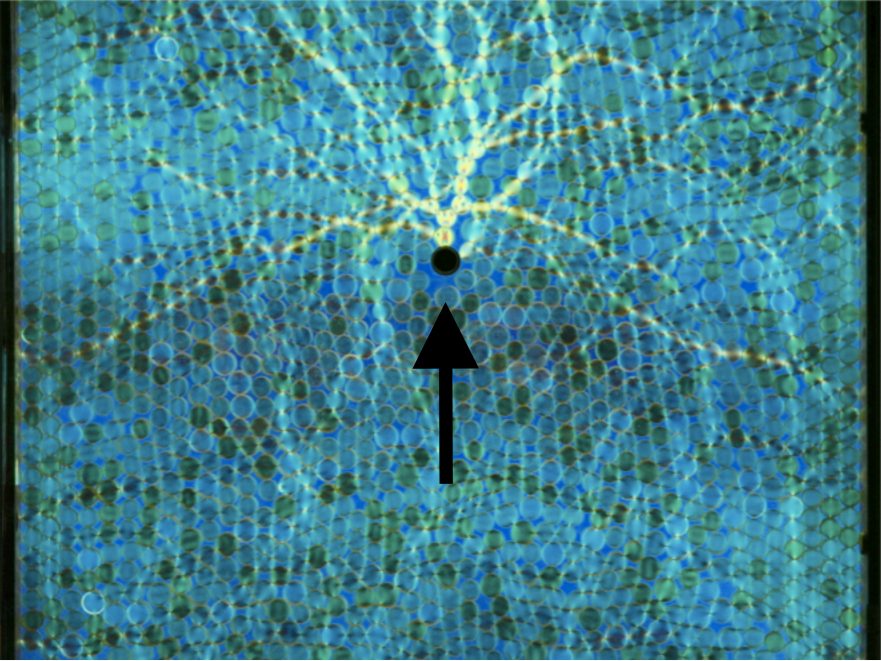}%
\caption{\label{fig:SolnQrt0} 
Left: the theoretical solution for $\sigma_{rr}$, given in (\ref{eq:stress1}), it `propagates' into the system along two symmetric characteristics that curve backwards.
Right: an experimental observation (by Prof. J. Zhang in Prof. R. P. Behringer's lab) of force chains curving backwards in an assembly of 2D photoelastic particles, when loaded by a local force in the direction indicated by the black arrow \cite{Zh16}. (Image courtesy of Prof. J. Zhang.)}
\end{center}
\end{figure}

The coupling between the characteristics makes it difficult to derive general analytic solutions and we resort below to insight derived from numerical solutions. 
However, it is instructive to analyse first the special case when $\Pi$ is diagonal. 
In this case, $\tilde{q}_{\theta\theta}<0$, $\tilde\lambda_2=-\tilde\lambda_1=\sqrt{-\tilde{q}_{\theta\theta}}$ and a straightforward calculation yields
\begin{equation}
\tilde{w}_1=\tilde{w}_2 = \frac{\sigma_{rr}(s=0)}{2}e^{-(\tilde{q}_{\theta\theta} + 1)s} \ ,
\label{eq:SolnQrt0}
\end{equation}
from which the stress is found exactly 
\begin{equation}
\left( \begin{array}{c} \sigma_{rr} \cr \sigma_{r\theta} \cr \sigma_{\theta\theta} \end{array} \right) = 
\left( \begin{array}{c} 1 \cr 0 \cr -\tilde{q}_{\theta\theta}\end{array} \right) \sigma_{rr}(s=0)e^{-(\tilde{q}_{\theta\theta} + 1)s}
\label{eq:stress1}
\end{equation}
An example of such a solution for $\sigma_{rr}$ is shown in figure \ref{fig:SolnQrt0}, when a narrow Gaussian $\sigma_{rr}$ stress source is applied at $(r,\theta)=(r_{0},0)$ (where $s=0$ for both characteristics). This solution exhibits several interesting features. 
One is the aforementioned curving of the stress paths. 
Another is flaring out of paths with $r$.
A third is a broadening of each path with $r$. 
A fourth is `leaking' of stress to the region between the paths, known as the cone of influence \cite{Geetal08}. This `leak' is also the cause of  attenuation of the stress along the paths.
Again, in solutions in rectangular coordinates, these phenomena cannot occur for spatially uniform fabric tensors, which would exhibit only straight path trajectories carrying constant stress. 

A fifth, and a spectacular feature, is that the stress chains can curve backwards! This underlines the difference between isostaticity and strain-based theories, where such a phenomenon cannot occur. In the isostatic medium, the 'back-bending forces' are balanced by the stress that leaks to the cone of influence. 
We have checked that the back bending of the stress chains does not lead to sign change of the stress anywhere in the entire system. This means that tensile forces do not develop anywhere in the system, which would have destabilised the structure of dry granular materials. Therefore, such solutions are physically viable. 
Indeed, back-bending force chains have been observed experimentally \cite{Zh16}, as can be seen in figure \ref{fig:SolnQrt0}. 
We can also predict the conditions for this phenomenon to be observed. At the point where bending back first occurs, the tangent to the path trajectory makes an angle of $\pi/2$ with the original orientation, which we designate as the $x$-axis. Writing this condition as $dx/dy=0$ and converting it to polar coordinates we get that this point corresponds to $\theta > \pi/4$ on the left branch and $\theta < -\pi/4$ on the right, or $\theta > \mid\pi/4\mid$. Using then (\ref{eq:ChainFlarea}), we find that the $i$th characteristic path starts bending back at a critical radius, $r_{ci}$, satisfying 
\begin{equation}
r_{c,i}=r_{0} e^{\pi/\left(4\tilde\lambda_i\right)}  \quad ; \quad i=1,2 \ .
\label{eq:BackForce}
\end{equation}
In other words, to observe stress chains bending back, the stresses along the paths should not attenuate to invisibility before the critical radius is reached.

\begin{figure}[h]
\begin{center}
\includegraphics[width=5cm]{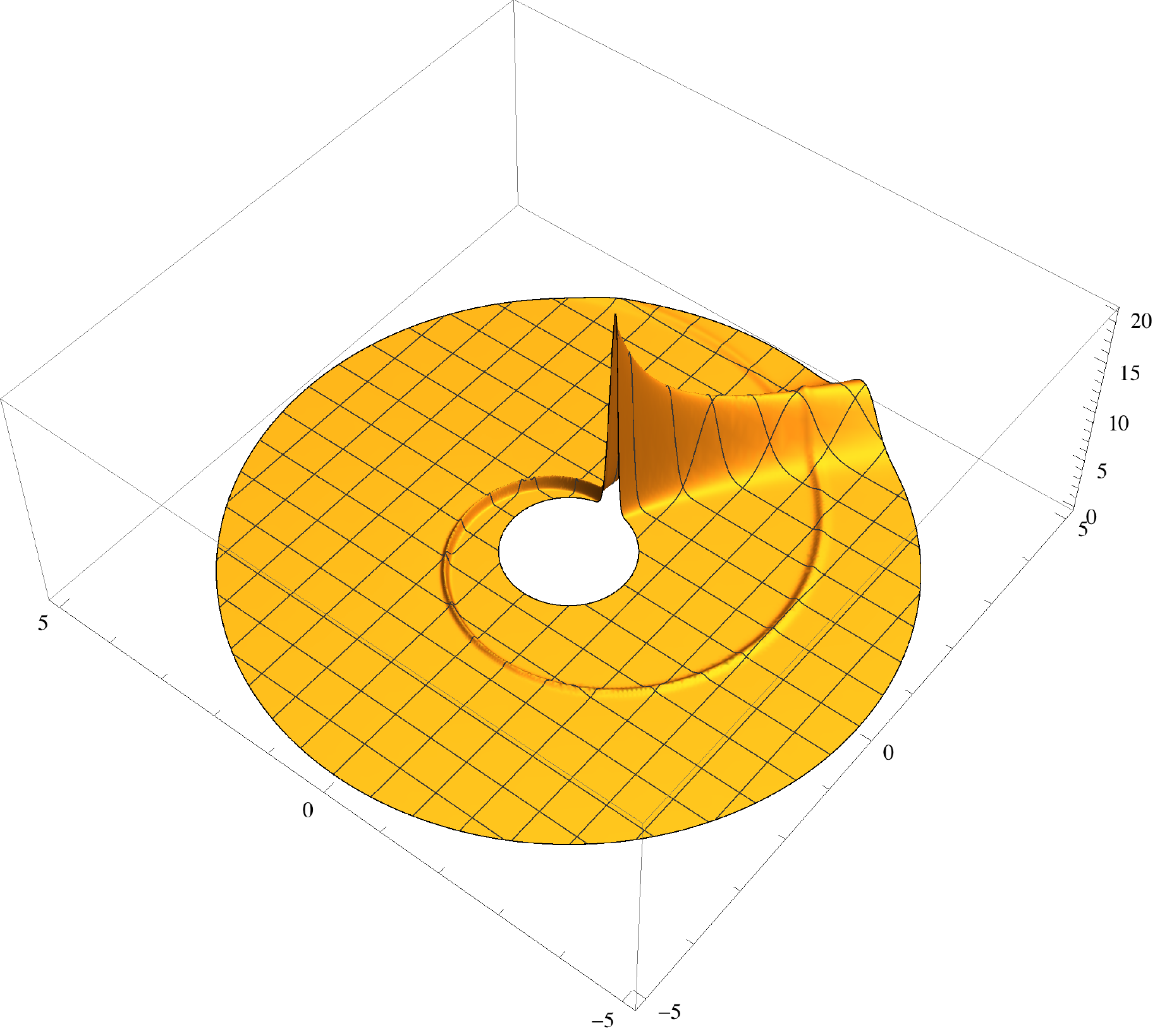}%
\includegraphics[width=5cm]{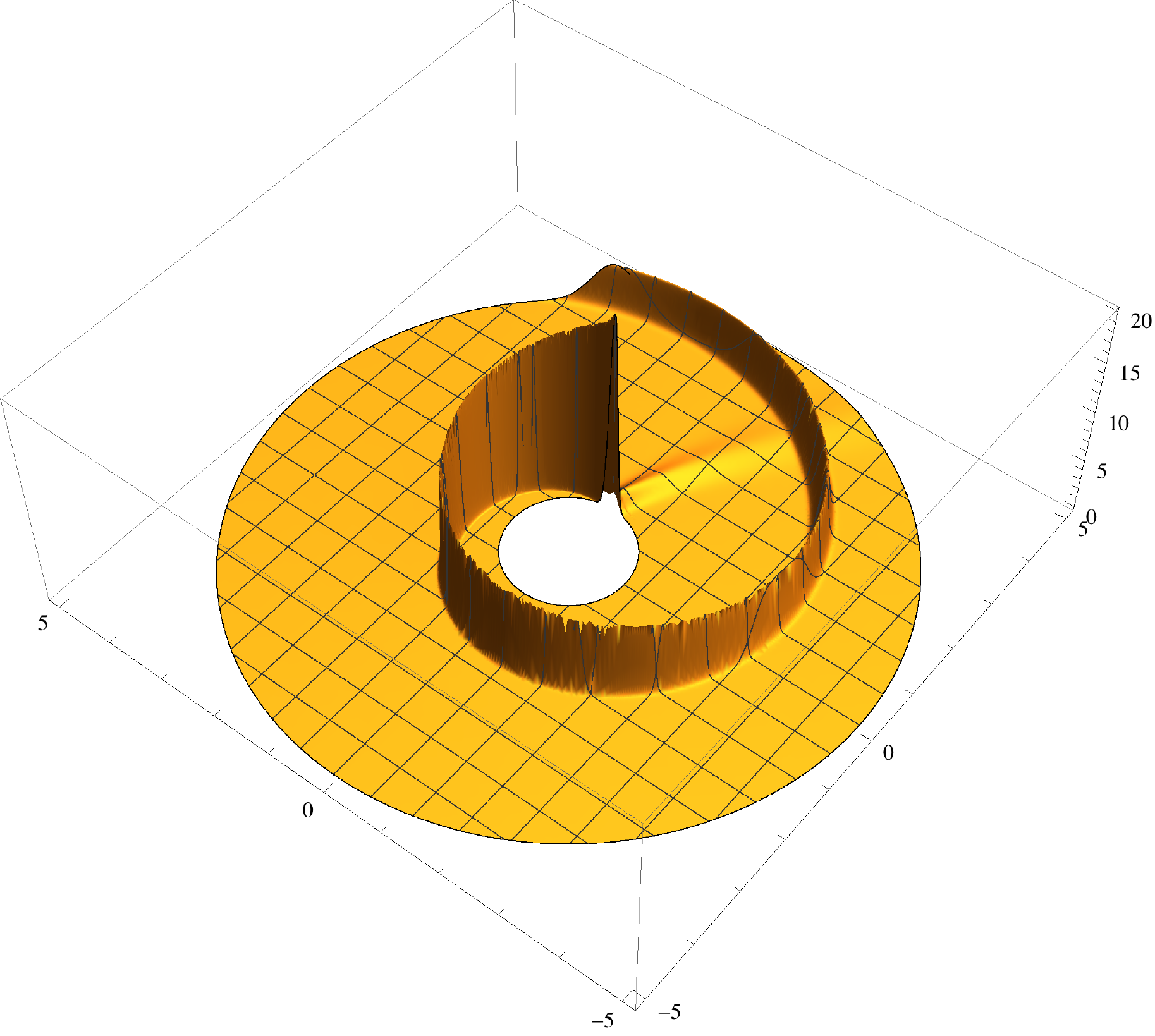}%
\caption{\label{fig:SolnQrt1} 
The solutions $\sigma_{rr}$ (left) and $\sigma_{\theta\theta}$ (right) for $\tilde{q}_{r\theta}=2$ and $\tilde{q}_{\theta\theta}=-1$. The two characteristic paths are asymmetric and one of them bends back before reaching the outer boundary, whilst the other does not. The stress is everywhere positive, indicating no development of tensile forces anywhere.}
\end{center}
\end{figure}

For more involved fabrics, when $\tilde{q}_{r\theta}\neq 0$, the characteristic paths may no longer be symmetric. We plot an extreme example of such a solution, obtained numerically for $\sigma_{rr}$ and $\sigma_{\theta\theta}$, in figure \ref{fig:SolnQrt1}. For this solution we used $\tilde{q}_{r\theta}=2$ and $\tilde{q}_{\theta\theta}=-1$. Note the bending back of one of the paths. The stress in this solution also remains compressive throughout the system.

\begin{figure}[h]
\begin{center}
\includegraphics[width=5cm]{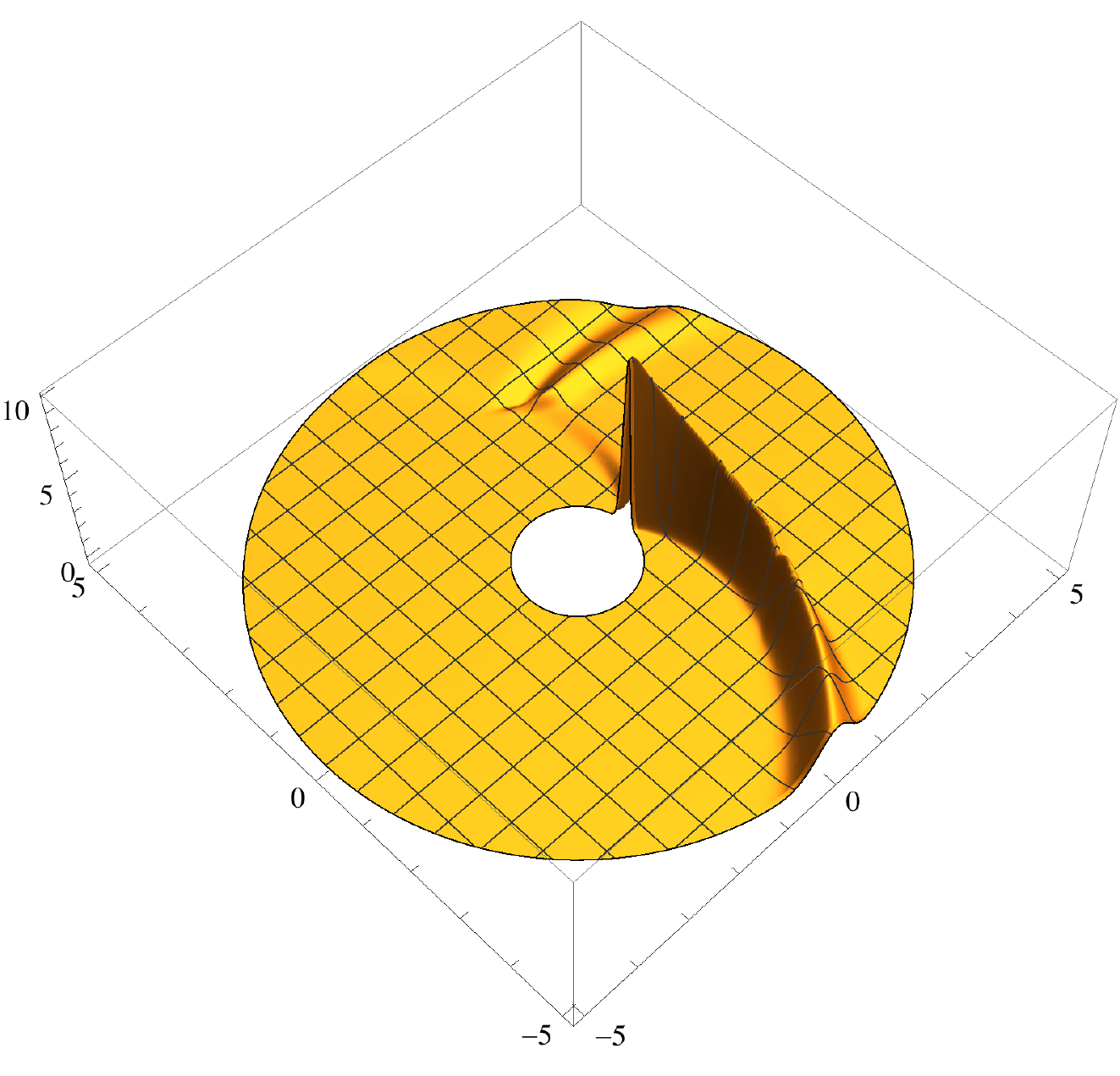}
\includegraphics[width=5cm]{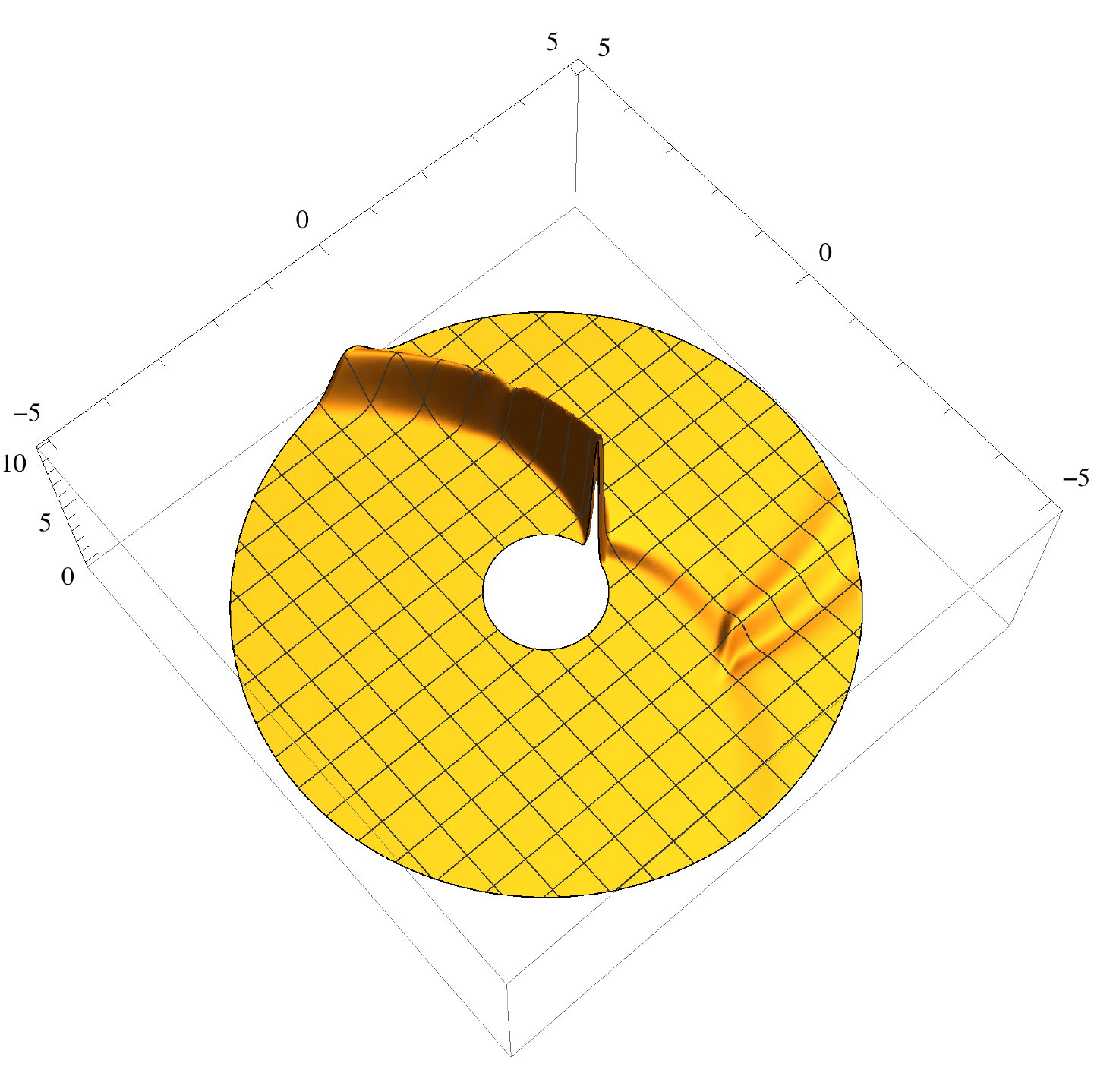}
\caption{\label{fig:Branch} 
The branching of the characteristic paths $\tilde{w}_1$ (left) and $\tilde{w}_2$ (right) when the fabric tensor has a local perturbation $\pi_{\theta\theta}=-1-exp\left[-100\left(r-3r_{in}\right)^2/r_0^2\right]$.}
\end{center}
\end{figure}

The full richness of the solutions in cylindrical geometries emerges for fabric tensors that vary spatially across the medium. Analytic solutions for such media are difficult to obtain and we resort to numerical solutions for insight. In the following, we keep to narrow Gaussian stress sources at the inner boundary. 
The example illustrated in figure \ref{fig:Branch} is of a localised perturbation to the fabric at  $r=3r_{0}$: $\tilde{q}_{\theta\theta}=-1-exp\left[-100\left(r-3r_{0}\right)^2/r_0^2\right]$. 
The perturbation gives rise to a clearly observed path branching. This is reminiscent of the branching observed in rectangular systems \cite{Geetal08}. 

\begin{figure}[h]
\begin{center}
\includegraphics[width=5cm]{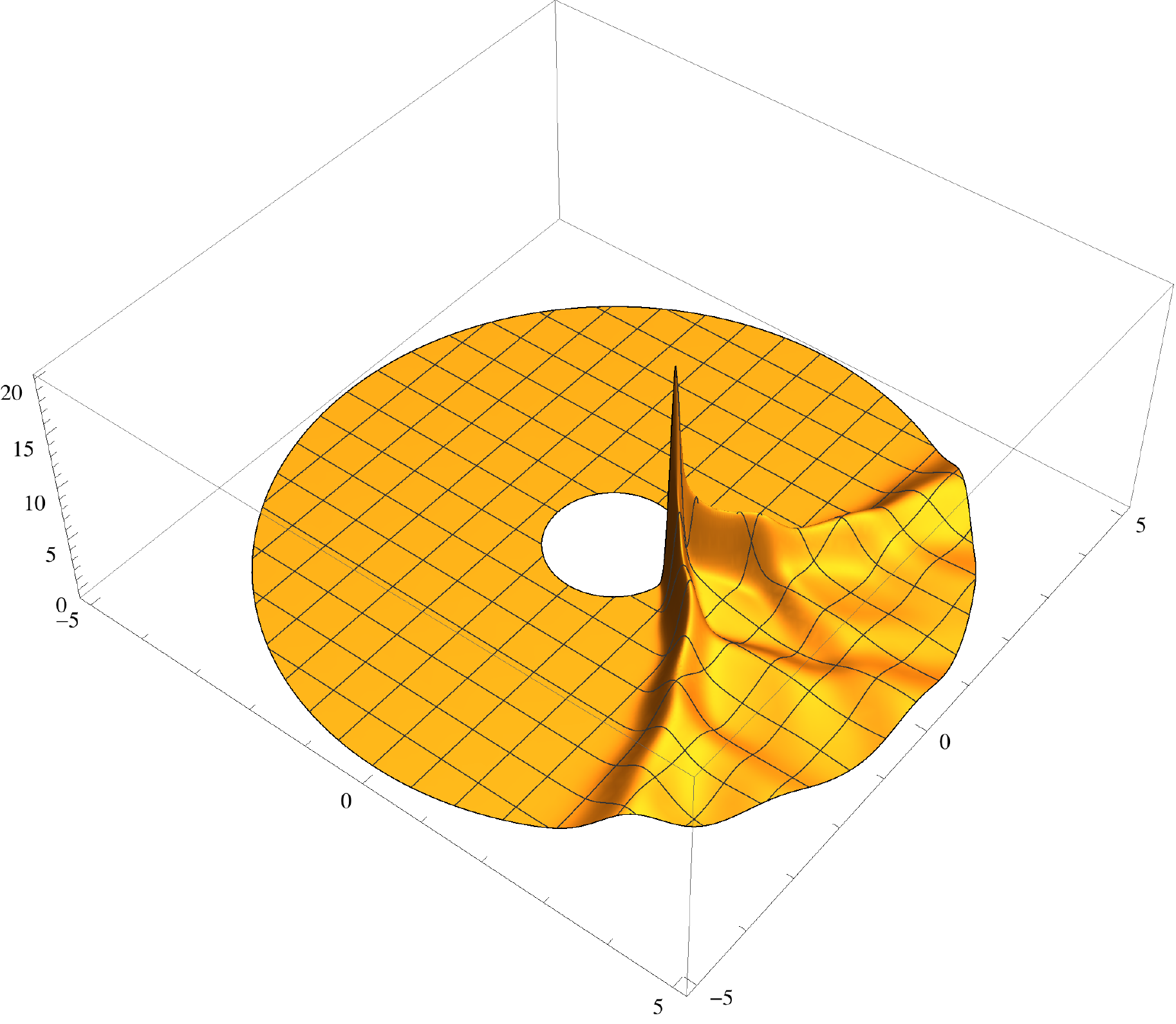}
\includegraphics[width=5cm]{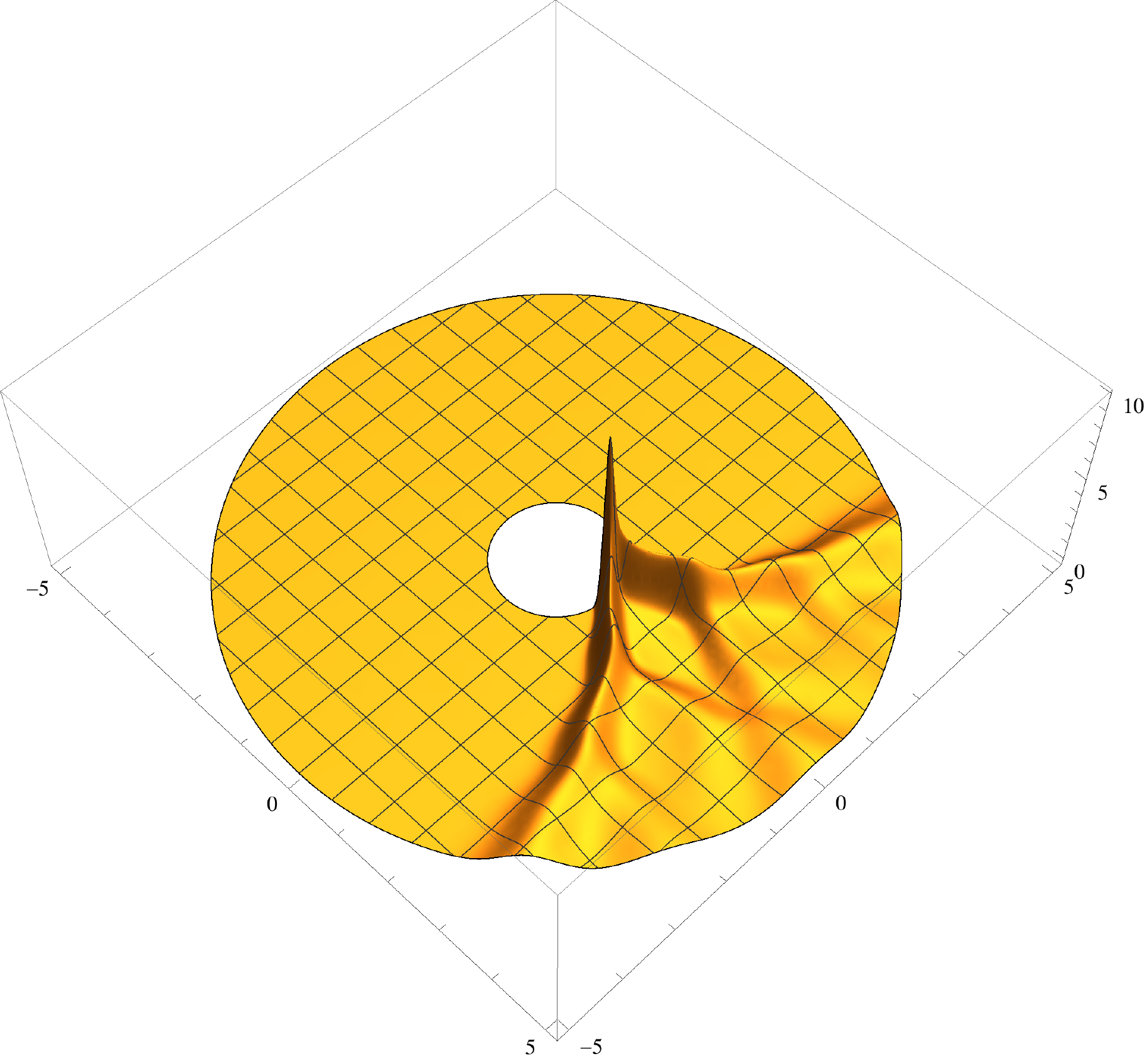}%
\caption{\label{fig:Periodic} 
The solutions $\sigma_{rr}$ (left) and $\sigma_{\theta\theta}$ (right) for $\tilde{q}_{r\theta}=0$ and a periodic $\tilde{q}_{\theta\theta}$. The characteristic paths are symmetric and leak stress into the cone of influence at periodic intervals. Outside this cone, $\sigma_{rr}=\sigma_{r\theta}=\sigma_{\theta\theta}=0$ identically. }
\end{center}
\end{figure}

The effects of spatial non-uniformity are nicely isolated and illustrated in the symmetric case and, for clear visualisation of the branching effect, we solve for the fabric tensor 
$\left(\tilde{q}_{rr}, \tilde{q}_{r\theta}, \tilde{q}_{\theta\theta}\right) = \left(1, 0, -0.2\cos(5r/r_0) - 0.3\right)$. The solutions for $\sigma_{rr}$ and $\sigma_{\theta\theta}$, plotted in figure \ref{fig:Periodic}, show clearly the periodic branch-like behaviour induced by the periodicity in the medium. The stress outside the cone of influence is identically zero whilst stress leaks from the main paths into the cone of influence at periodic intervals, where the gradient of the fabric tensor are largest. The leakage is along secondary characteristic paths.


\section{Experimental support}
\label{sec:5}
Several experiments support these results. Flaring of force chains has been observed, e.g. in figure 7 of \cite{Geetal03} and in our figure \ref{fig:SolnQrt0} \cite{Zh16}. Back-bending forces have also been observed, as figure \ref{fig:SolnQrt0} shows.  
Another intriguing and potentially related experimental observation has been reported in \cite{BoRe09}. The experiments consisted of shearing sand-filled Couette cells by a very slowly rotating inner boundary and observing formation of patterns of Mandala-like slip lines in the medium. The slip lines were narrow, well defined and appeared in pairs, flaring out almost exactly symmetrically from points along the inner boundary (their figure 4). Moreover, the slip lines pass through one another with very little interaction, if at all. 
The pattern and shapes of these slip lines are identical to the stress chain solutions derived here and it is tempting to relate the two. Our interpretation of their observations is that, on the verge of slipping, their medium is almost perfectly isostatic and the low shear rate makes it possible for the medium to remain very close to this state, making possible the appearance of our solutions. The slow shear generates distinct stress sources along the inner cylinder, which give rise to characteristic paths in the material, following the analysis presented here. The slip lines then form at the zones of highest shear stress. The slow rotation of the inner axis breaks somewhat the left-right symmetry, which is why one characteristic family is more evident than the other.
\begin{figure}[h]
\begin{center}
\includegraphics[width=3.9cm]{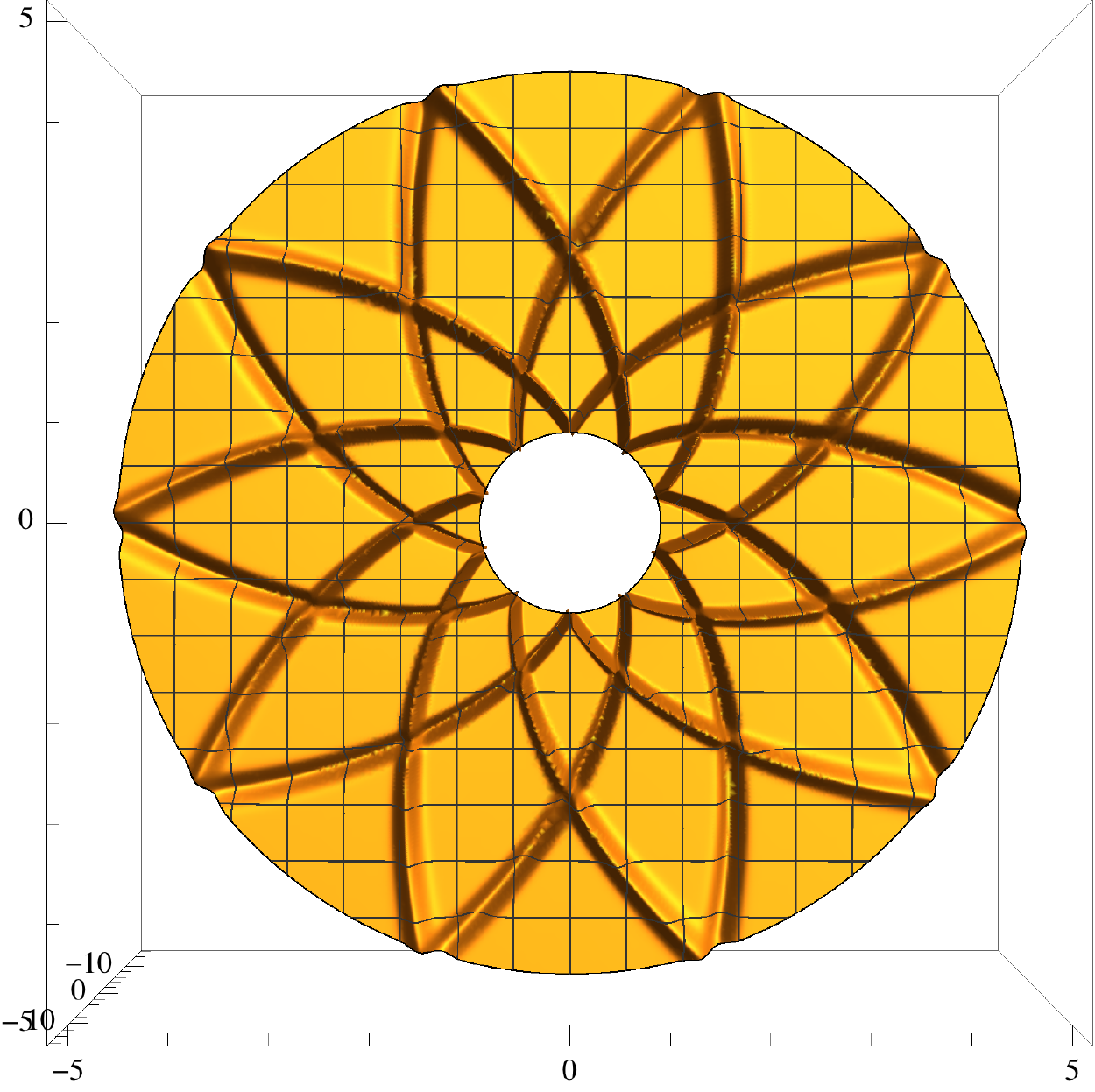}
\includegraphics[width=3.9cm]{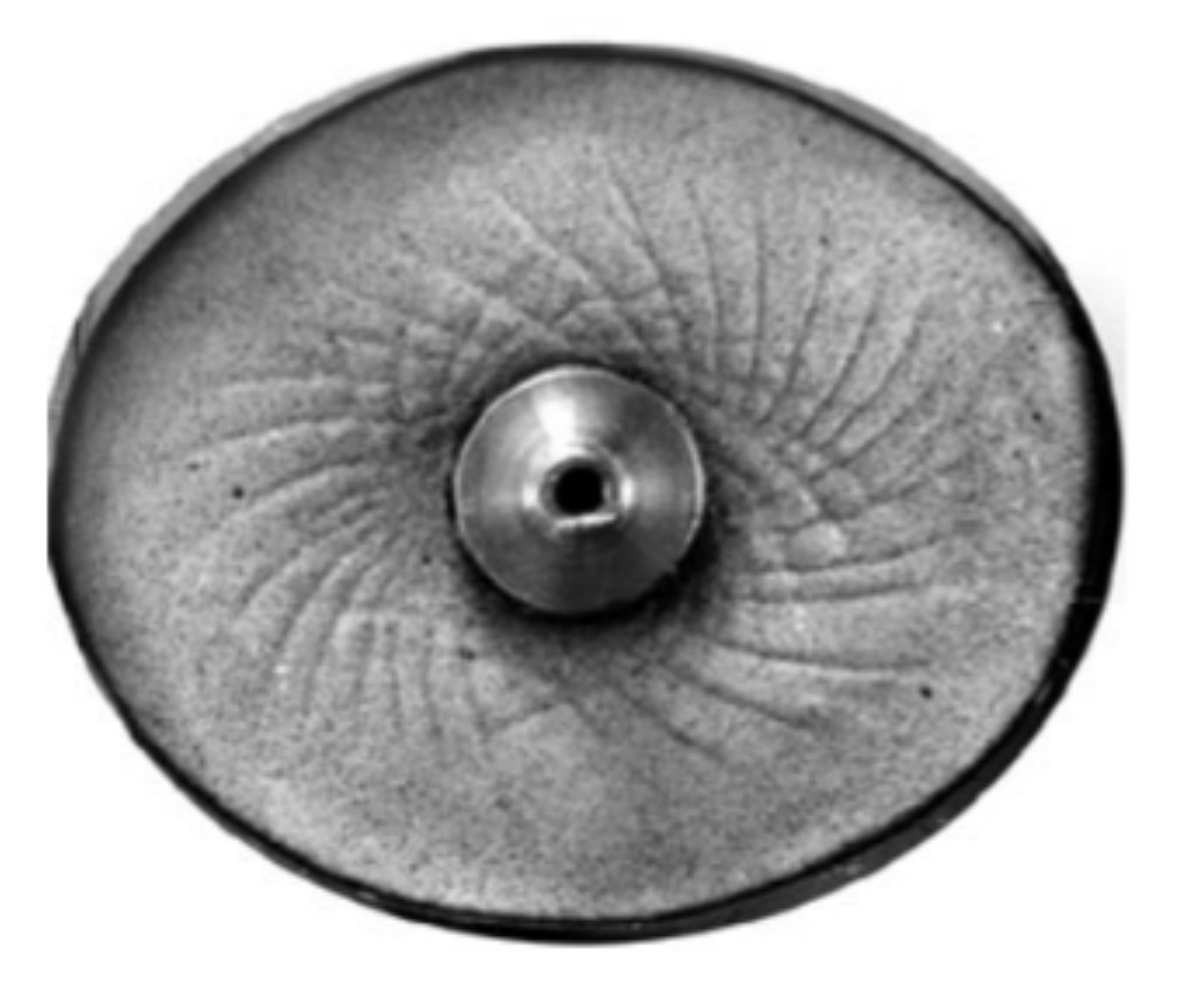}%
\includegraphics[width=3.9cm]{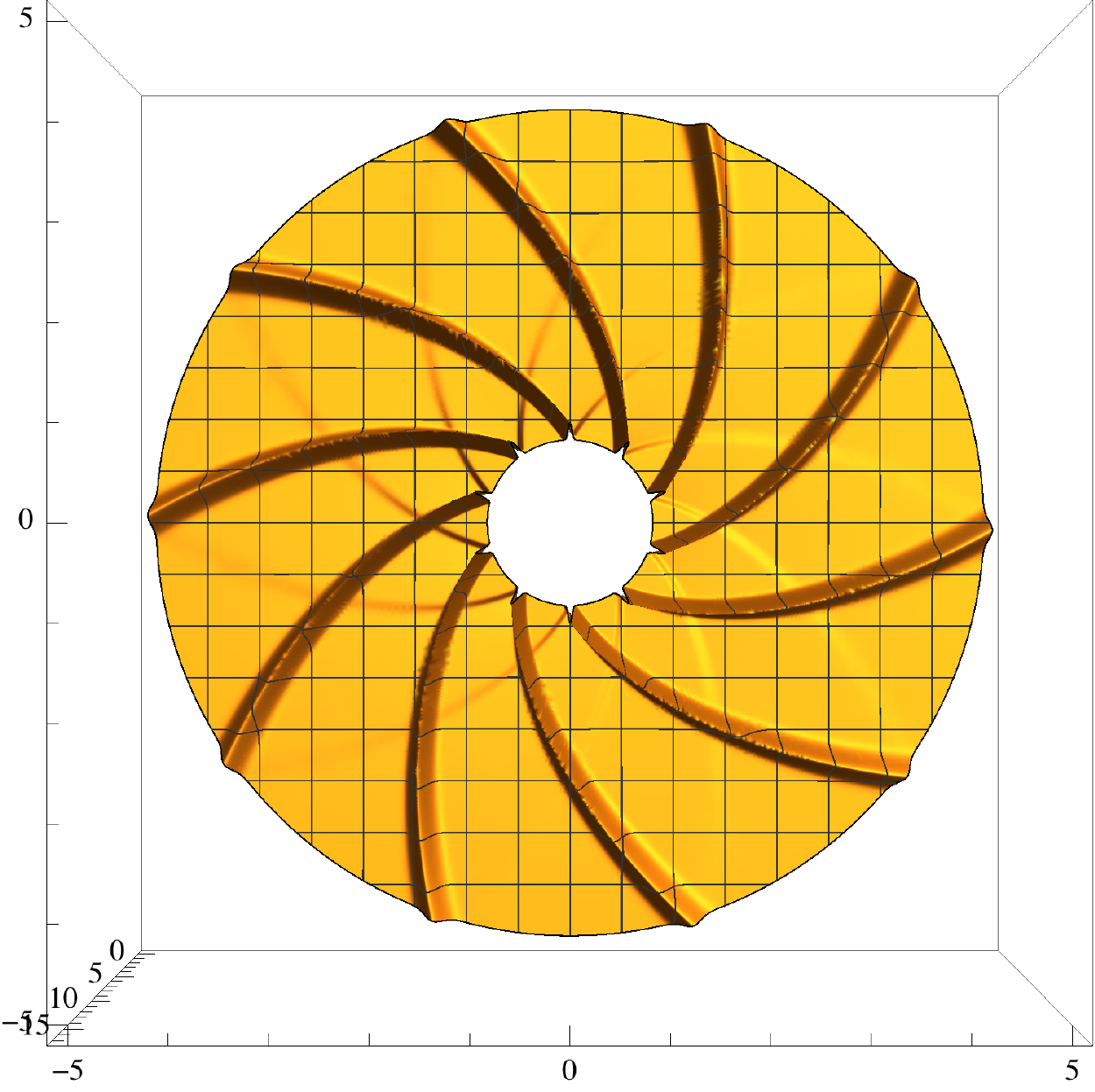}%
\caption{\label{fig:CylShearChains} 
The $\sigma_{r\theta}$-chain solutions (left) for the fabric tensor measured from figure 4 of \cite{BoRe09}. For a clear comparison with the experimentally observed Mandala-like slip lines (centre, courtesy of Bobryakov and Revuzhenko \cite{BoRe09}), only ten chain pairs are shown. The rotation in the experiment breaks the left-right symmetry and gives slight preference to one family of characteristics over the other (right).  The chains geometries compare well to the experiment despite the rough estimate from their figure.}
\end{center}
\end{figure}

We can further use their figure 4 and data to obtain information about the structural constitutive properties of the granular medium in their experiments. Assuming that the structure can be approximated by a spatially uniform fabric tensor, the fact that there is very little interaction between the slip lines suggests that $\pi_{r\theta}=0$. This also accounts for the near-symmetry of the slip line pairs. We can also estimate, from the figure and the reported data, an approximate relation between the flaring out angle and radius along every slip line. Using this estimate in our equation (\ref{eq:ChainFlarea}) we deduce that their effective fabric tensor satisfies $\tilde\lambda_{1,2} \approx \pm \frac{\pi}{6\ln{2.5}}$ and hence that $\tilde{q}_{\theta\theta}=-\tilde\lambda_{i}^2 = -0.33\pm0.06$, where the error stems from the uncertainty of measurements from their figure 4. 
In figure \ref{fig:CylShearChains}, we plot the shear stress chains, $\sigma_{r\theta}$, for this fabric tensor. For clarity, we show only ten pairs, as the experimentally produced twenty plus pairs would clutter the figure. We also plot the left hand side characteristic paths family on its own for clear comparison. The strong similarity between the experimental and theoretically derived chain geometries supports both our analysis, as well as our interpretation of their observations. 

It should be noted, however, that this Mandala-like pattern can only arise when the fabric tensor is very close to uniform across the system and $\pi_{r\theta}=0$. It is very likely that this is the case in the experiment because of the careful initial preparation of the material and the very slow shear rate. More general fabric tensors are expected to be more disordered, having have local gradients, which would give rise to secondary disordered stress chains.

\section{Conclusion and discussion}
\label{sec:6}
To conclude, we analysed the isostaticity stress field equations in polar coordinates. We derived the equivalence relation between the constitutive fabric tensors in cartesian and polar coordinates and the constraint that the components of the latter must satisfy for the equations to be hyperbolic and yield stress chain solutions. 
The stress equations were analysed and an explicit formula has been obtained for the flaring out of the stress chains and their broadening away from the inner boundary. This relation was then used to show that stress chains can bend back and exert force components in a direction opposite to the original loading! This striking phenomenon, which is impossible in strain-based theories, such as elasticity, is a fingerprint of the arching effect. This  phenomenon is not dissimilar to chains of dominos, which can be made to fall, and thus have momentum, in opposite direction to the initial domino. 

We emphasise that this analysis is significant beyond the relevance to cylindrical geometries and Couette cells. It is often the case that an external load is localised almost at the particle level, say due to grains being pressed by the boundaries more than their neighbours. Since the size of the system is normally much larger than the localised length scale,  the stress field around such a source has locally unavoidably a cylindrical symmetry. Consequently, the stress field near a source is better described by equations (\ref{eq:Balance3})-(\ref{eq:ConstEq2}) then by (\ref{eq:Balance1})-(\ref{eq:ConstEq}). This suggests that some of the effects we derived here may also be observed in experiments other than in Couette cells. 

It is interesting that the flaring out phenomenon does not arise in most of the solutions in the literature, where rectangular coordinates were used and straight stress chains were found for uniform fabric tensors. We believe that the reason for the apparent discrepancy between those and the solutions studied here is that we constrain our fabric tensors to have a negative determinant {\it for all azimuthal angles $\theta$}, a constraint that is missing from those initial studies. Moreover, in view of the above discussion, not imposing this constraint leads to regions in the plane where the fabric tensors, used in the literature, may have positive determinants, as shown by our eq. (8). In reality, this means that stress chains may disperse when incident on a region with a positive determinant of the fabric tensor, where the equations become locally elliptic.

Numerical solutions were then derived for several uniform fabric tensors in annuli, supporting the analytical results and demonstrating that stress chains indeed flare out, broaden and bend back. These solutions also demonstrated the further rich behaviour in cylindrical systems: chains leak stress into the region between the chain pairs - the cone of influence. 
That these phenomena occur even for perfectly uniform media, when $\nabla{\pi_{ij}}=0$, is in stark contrast with solutions in rectangular geometries, where such fabric tensors can only give rise to constant stress along straight stress paths and zero stress elsewhere. 

We then studied several numerical solutions for spatially varying media and showed that large gradients in the local fabric tensor lead to stress chain branching, a phenomenon seen also in rectangular geometries \cite{Geetal08}. This branching is in effect a strong `leak' from a localised region on a specific path, whose trajectory is along a conjugate secondary characteristic path into the cone of influence. 
Effects of general spatially varying structures were also illustrated by solving for a fabric tensor with a component periodic in the azimuthal angle. The periodicity in the fabric induces periodic  leaking stresses from the main paths, again via secondary paths into the cone of influence. A similar phenomenon would be observed for periodicity in the $r$-direction. 

Finally, we pointed out experimental observations that not only support our results but also can be explained afresh in view of them. To the best of our knowledge, back-bending forces, although observed in experiments, have been neither discussed nor studied in the literature. It would be interesting to study this phenomenon in more detail in light of the predictions made here. Our results also suggest a new {\it quantitative} interpretation of the shearing experiment in \cite{BoRe09}.

This work can be extended in a number of directions. 
Theoretically, the next natural step is to obtain solutions for nonuniform fabric tensors and test how the statistics of their local gradients affect the statistics of the main stress chains, as well as those of the secondary and tertiary branching ones. 
Following the discussion above, concerning the differences between the solutions in rectangular and polar coordinates, another direction to explore is a more detailed understanding of the behaviour of stress chains for the fabric tensors, conjectured in the literature for rectangular coordinates, once used as input for our equations here. In particular, it would be interesting to observe how straight stress chains \cite{Edinburgh,Geetal08} disperse upon entering an elliptic region of the fabric tensor. 
Another natural extension is to three dimensional systems, but this extension has to wait until a proper such theory is constructed even for rectangular coordinates. 
These theoretical directions are being taken currently in our group. 
We are also looking forward to real and numerical experiments to test our theory, in particular of very slow shear in Couette cells. For example, related simulations \cite{Laetal00} found an inclination of the principal stress direction relative to the radial direction, which initially increases with radius and then saturates to a constant of about $\pi/4$. Our theory predicts that this inclination, which is the direction of the stress chain, would increase continuously. Those experiments, while relevant, are inconclusive as a test. Firstly, the increase found there is over a 2-particles thick shear band, which cannot be considered a continuum. Secondly, their observed saturation to a constant angle is probably due to the increasing connectivity away from the shear band, which takes the medium away from the isostatic state, where the theory is valid. Thirdly, our theory predicts two such principal stresses emanating from the inner cylinder, while they observe only one. This could also be a result of the broken symmetry by the shear. Nevertheless, in view of our results, it would be useful to modify such experiments accordingly to provide a rigorous of our solutions, e.g. by generating wider isostatic-like shear bands.




\end{document}